\documentclass[12pt,a4paper]{article}
\usepackage{palatino,cite,epsfig,epsf,amsmath,amssymb,color}
\usepackage{graphics,graphicx,rotating,caption2,colordvi}
\usepackage{hyperref}
\usepackage{siunitx}
\usepackage{mathptmx}
\usepackage{graphicx} 
\usepackage{soul} 
\usepackage{float}

\usepackage{amsmath}
\usepackage{amsfonts}
\usepackage{amssymb}
\usepackage[a4paper, left=25mm, right=25mm, top=25mm, bottom=20mm]{geometry}

\begin{document}

\begin{titlepage}

{\flushright{
        \begin{minipage}{2.5cm}
         DESY-23-138
        \end{minipage}        }

}

\begin{center}
{\LARGE\bf
R\&D for Positron Sources at High-Energy Lepton Colliders
\footnote{Talk presented at the International Workshop on Future Linear Colliders (LCWS2023), 15-19 May 2023. C23-05-15.3.}}
\vskip 1.0cm
{\large Gudrid Moortgat-Pick$^{a,b}$,
Sabine Riemann$^{c}$, 
Peter Sievers$^{d}$,
Carmen Tenholt$^{b}$}
        \vspace*{8mm}\\
{\sl\small 
${}^a$ University of Hamburg, Luruper Chaussee 149, 22761 Hamburg, Germany \\
${}^b$ Deutsches Elektronen-Synchrotron DESY,  Notkestr.\ 85, 22607 Hamburg, Germany\\
${}^c$ Deutsches Elektronen-Synchrotron DESY, Platanenallee 6, 15738 Zeuthen, Germany\\
${}^d$ CERN CH 1211 Geneva 23, Switzerland}
\end{center}

\begin{abstract}
Several designs for high-energy Lepton Colliders serving as Higgs factories but extendable to higher energies up to the TeV range are under discussion. The most mature design is the International Linear Collider (ILC), but also the Compact Linear Collider (CLIC) as well as the new concept  of a Hybrid Asymmetric Linear Higgs Factory (HALHF) have a large physics potential. The first energy stage with $\sqrt{s}=250$~GeV requires high luminosity and polarized beams  and imposes an effort for all positron source designs at high-energy colliders. In the baseline design of the ILC, an undulator-based source is foreseen for the positron source in order to match the physics requirements. In this contribution an overview is given about the undulator-based source, the target tests, the rotating target wheel design, as well as the pulsed solenoid and the new technology development of plasma lenses as optic matching devices. 
\end{abstract}

\end{titlepage}

\section{Motivation}
A large amount of data has been collected from the Large Hadron Collider (LHC) and highly involved and precise analyses methods have been used, but no further discovery has been confirmed ---although several excesses (e.g. consistent with a light scalar around 95 GeV)  have been detected--- and a possible window to new physics is still unknown. Nevertheless, there still exists a strong motivation for new physics (NP) beyond the Standard Model (SM), as, for instance, the quest for dark matter  candidates, for explanation of the baryon-antibaryon asymmetry etc. 
Therefore further additional tools at future Lepton Colliders complementary to the High-Luminosity LHC (HL-LHC) analyses are required 
in order to identify promising windows to NP.
The ILC~\cite{Adolphsen:2013kya,ilc-rdr,BCD}  as well as the multi-TeV high-energy collider design CLIC~\cite{Aicheler:2012bya} provide 
polarized beams at high intensity as well as at high energy. Challenging is the production of the high-intense positron beams.
The ILC uses an undulator-based positron source in the baseline design~\cite{Adolphsen:2013kya,ilc-rdr,BCD} that produces a polarized positron beam
already for the first stage~\cite{Ushakov:2018wlt}, i.e. an initial energy of $\sqrt{s}=250$~GeV \cite{Fujii:2018mli}, but is also applicable for higher centre-of-mass energies~\cite{Mikhailichenko:2016iun}.
With simultaneously polarized beams and high luminosity\cite{Moortgat-Pick:2005jsx}, the physics potential of the ILC is optimized and well prepared for high precision physics as well as for new discoveries~\cite{Moortgat-Pick:2015lbx,AguilarSaavedra:2001rg}. There exists also the possibility to use an undulator-based positron source for CLIC~\cite{Liu:2010bb}. The implementation of polarized beams at the HALHF concept~\cite{Foster:2023bmq}
is currently under development.
Alternative designs for generating polarized positrons via Laser-Compton backscattering have been proposed as well but are currently not pursued~\cite{Kuriki:2008zz,POSIPOL:2006hts}.

It has already been shown that the physics precision requirements can not be fulfilled if only polarized electrons were available since
in that case the systematic uncertainties get too large, see \cite{Robert-Thesis,Karl:2017xra,Aihara:2019gcq}. 
But also new physics searches might rely substantially on the availability of both beams polarized, see, for instance \cite{Cheng-Li-pol}, where CP-violating
Higgs couplings might be detectable already at $\sqrt{s}=250$~GeV for transversely-polarized beams.
Therefore the use of simultaneously-polarized beams is crucial for new physics searches at future lepton colliders.

\subsection{Add-on's when using simultaneously polarized $e^-$ and $e^+$ beams}
Already at the first energy stage of the ILC with a centre-of-mass energy of $\sqrt{s}=250$~GeV (ILC250) a polarized positron beam with $P(e^+)=\pm 30\%$ would be available, upgradable to $P(e^+)=\pm 40\%$ ($P(e^+)=\pm 60\%$ after implementation of the photon collimator) 
from $\sqrt{s}=350$~GeV onwards, see \cite{ Adolphsen:2013kya,Riemann:2020ytg} for more technical details. The implementation of spin rotators and -flippers before the damping ring is  required to switch helicities of both 
$e^{\pm}$ beams with similar frequency to exploit the increase in effective luminosity\cite{Riemann:2012zza}. Furthermore, the spin rotators
allow both the use of longitudinally- as well as transversely-polarized beams for physics interactions.

\begin{figure}
	\includegraphics[scale=.48]{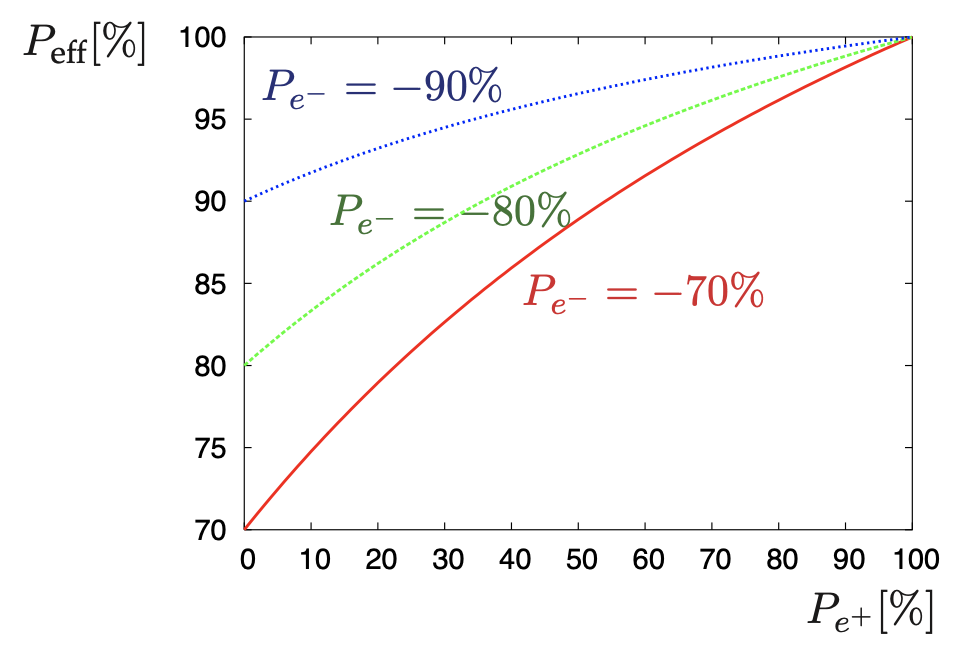}
	\includegraphics[scale=.48]{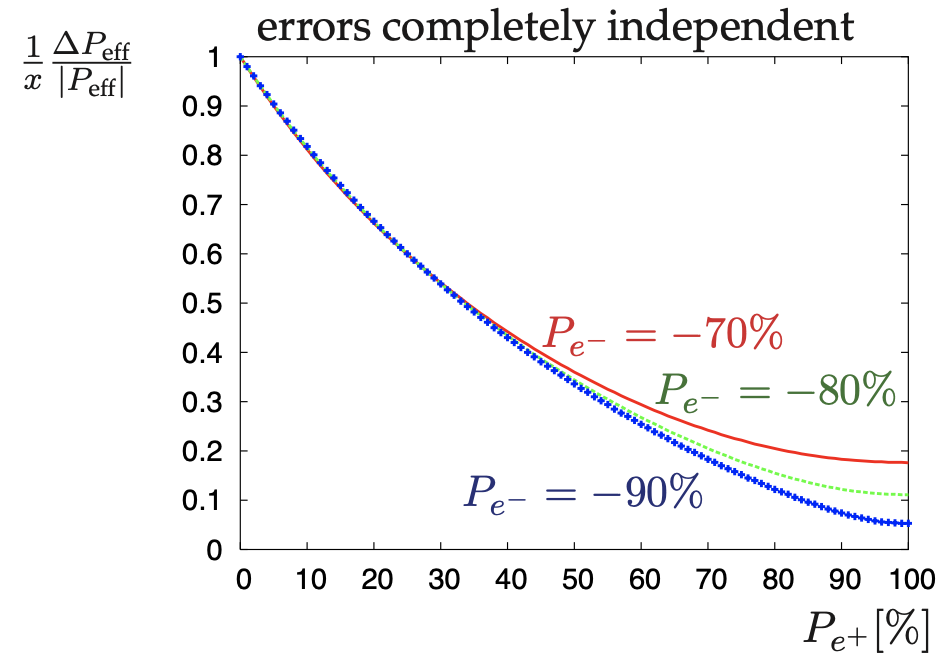}	 
\caption{Left panel: Effective polarization vs.\ positron 
beam polarization.Relative uncertainty on 
the effective polarization. Right panel: 
$\Delta P_{\rm eff}/ |P_{\rm eff}|\sim \Delta A_{\rm LR}/A_{\rm LR}$, 
normalized to the relative polarimeter
precision $x=\Delta P_{e^-}/P_{e^-}=\Delta P_{e^+}/P_{e^+}$ 
\cite{Moortgat-Pick:2005jsx}.
\label{fig:alr}}
\end{figure}

The availability of simultaneously polarized $e^-$ as well as $e^+$ beams leads to several advantages ~\cite{Moortgat-Pick:2005jsx}, since 
positron polarization plays the key role to obtain the following benefits and has a large impact on new physics contributions in Higgs, $WW$, $Z$ and $t$ physics, starting already at the first energy stage of $\sqrt{s}=250$~GeV!
\begin{itemize}
\item Better statistics: higher effective polarization $P_{\rm eff}:=(P_{e^-}-P_{e^+})/(1-P_{e^-}P_{e^+})$, see Fig.\ref{fig:alr} (left panel),
 leads to higher rates and suppresses background processes; one expects at the ILC250 
$P_{\rm eff}=(\mp 90\%,\pm 30\%)=94\%$ and at the upgraded ILC500 
$P_{\rm eff}=(\mp 90\%,\pm 60\%)=97\%$.  
 \item Higher collision rates: higher effective luminosity  ${\cal L}_{\rm eff}/{\cal L}:=\frac{1}{2}(1-P_{e^-}P_{e^+})$ leads to a higher fraction of actual collisions and 
 offers therefore to achieve a specific number of event in less running time;
one gets at the ILC250 an enhancement in the effective luminosity by a factor of $1.3$ and at the upgraded ILC500 
by a factor of $1.5$ compared to the case with only polarized electron beams.  
The availability of both beams polarized reduces therefore the required running time by one third. 
\item Higher number of independent and new observables: four different data sets, opposite-site and like-sign polarization configurations can be evaluated and a larger set of polarization asymmetries can be exploited.
 \item But $P_{e^+}$ is not only required to improve significantly the statistics and saves running time but in particular also for getting systematic errors under control. Suitable observables are, for instance, left-right-asymmetries $A_{\rm LR}$.
 
Applying polarized beams in general requires to measure the polarization of both beams as accurate as possible~\cite{Moortgat-Pick:2005jsx,Moortgat-Pick:2015lbx}. Using Compton polarimeters up- and downstream an polarization uncertainty of $\Delta P/P=0.25\%$ is expected.
 A conservative lower bound for the resulting uncertainty in deriving $A_{\rm LR}$,  $\Delta P/P\sim \Delta A_{\rm LR}/A_{\rm LR}$ is shown in Fig.\ref{fig:alr} (right panel), where only the uncertainty in the polarization measurement
has been taken into account: the uncertainty $\Delta A_{\rm LR}/A_{\rm LR}$ 
 is reduced by a factor 2 at ILC250 and by a factor $>3$ at the ILC500 when providing polarized positrons.
\item Deriving the polarization via in-situ measurements of physics processes, for instance $W^+W^-$ production with simultaneously-polarized $e^{\pm}$ beams, the uncertainty can be reduced to $\Delta P/P=0.10\%$\cite{Karl:2017xra,Moortgat-Pick:2005jsx}.
\item Concerning vector- and axialvector-interactions, transversely-polarized beams --due to the negligible electron mass--- can only be exploited if both $e^{\pm}$ beams are simultaneously polarized. Already at the first energy stage $\sqrt{s}=250$~GeV, this option can very substantial for Higgs physics.
As shown in \cite{Cheng-Li-pol}, CP-violating couplings in the Higgs sector, for instance in the $HZZ$ coupling, can be probed with an unprecedented precision, if transversely-polarized beams as well as high luminosity (i.e. ${\cal L}^{-1}\ge 2000$~fb) are available. Since the polarization enters bilinearly $\sim P^{\rm T}_{e^-} P^{\rm T}_{e^+}$ highest possible polarization degrees are advantageous.
\end{itemize}


\section{Undulator-based positron source}
\begin{figure}[H]
	\centering 
	\includegraphics[scale=.5]{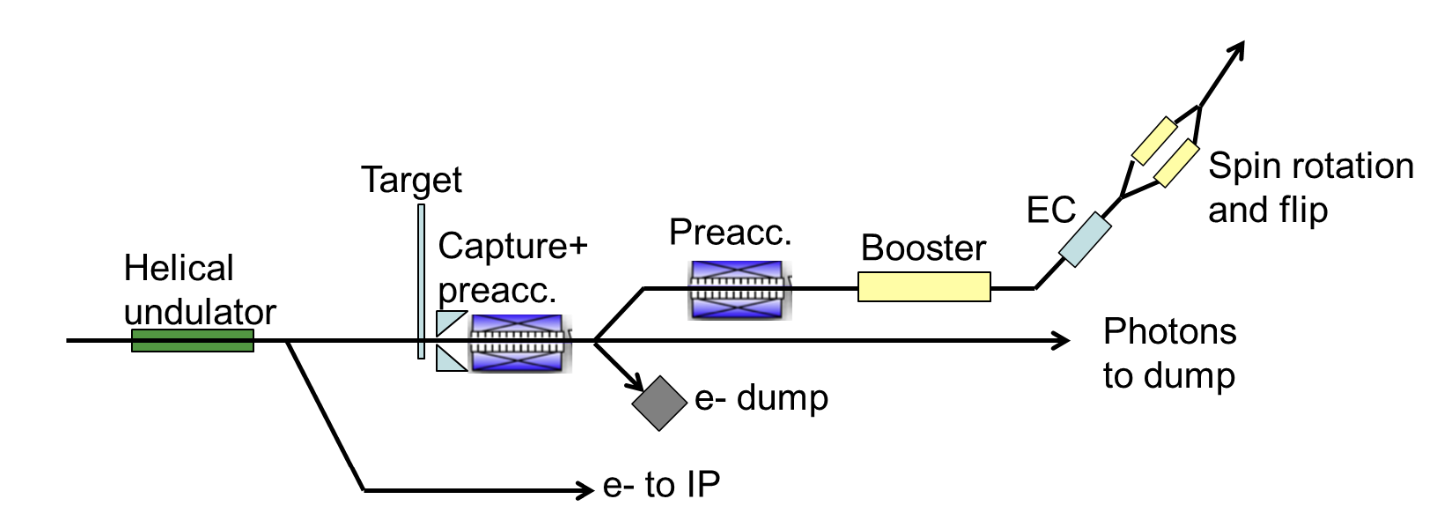} 
    \caption{Schematic layout of the ILC positron source \cite{Adolphsen:2013kya}.}
    \label{fig:positron-source-layout}
\end{figure}

The baseline design for the ILC positron source~\cite{Adolphsen:2013kya} uses a long  helical undulator passed by the high energy electron beam to create  an intense circularly polarized  photon beam. A proof-of-principle experiment of such a scheme for polarizing positrons has  been done with the E166 experiment\cite{E166} at SLAC. The photon beam hits a thin conversion target to produce electron-positron pairs. The target is designed as wheel of 1\,m diameter and  spinning with 2000 rounds per minute in vacuum avoiding overheating of the target material, i.e.\ Ti6Al4V. More details on  performed target material tests can be found in~\cite{Tim-proc}. Since the average energy  deposition in the target is only a few kW, cooling by thermal radiation is feasible. 
Inside the target electroweak interactions between atoms and photons with energies above a lower limit result in pair production of longitudinally polarized electrons and positrons. Only the positrons are kept; the parameters of these still divergent positrons are required to be matched to the acceptance requirements of the downstream damping ring. This matching can be achieved with the optical matching device (OMD), which is currently foreseen to be a pulsed solenoid, see  section~\ref{ref:OMD}. But also new technologies are currently been tested, e.g. a plasma lens prototype, to serve as OMD, 
see \cite{Niclas-proc,Manuel-proc}.
The challenge is that the ILC requires $1.3\times 10^{14}$ positrons per second at the interaction point (IP), that's a factor 100 more positrons than at the SLC.
Therefore a yield of $Y=1.5 e^+/e^-$ at the damping ring has to be fulfilled.
The OMD is followed by the capture RF cavity, which accelerates the positron bunch to $\SI{125}{\MeV}$. Further downstream elements before the damping ring are the electron and photon dump, SCRF booster, spin rotation solenoid, energy compression structure, 
cf.\  Fig.~\ref{fig:positron-source-layout}~\cite{Adolphsen:2013kya}.

Since the TDR great experience with long (planar) undulators has been achieved at the XFEL accelerator, resulting in extremely stable operation~\cite{Decking-proc}:
no radiation damage to the undulators has been occurred as well as no energy loss due to particle loss has been measured.
A beam alignment up to 10-20 microns for 200 m (= undulator length) has been achieved and during beam operation the beam trajectory is controlled better than 3 micron due to both slow and fast feedback systems~\cite{Decking-proc}.  Since the tolerances for the ILC undulator are even more relaxed than for
XFEL no operation and alignment issues are expected for the ILC undulator.

\subsection{Current undulator simulations w/wo masks}
Detailed simulations for the undulator-based source have been made including possible fields  and alignment errors.
It is already known that if the beam spot size slightly increases (e.g. by $\sim 45\%$), the yield also slightly decreases (by $3\%$)~\cite{Andriy-talk}.
In the recent study\cite{Khaled}  such non-ideal helical undulator calculations have been  made for the energy stages $\sqrt{s}=250$, $350$, $500$~GeV as well as for the GigaZ option.

\begin{figure}[H]
	\centering
	\includegraphics[scale=.4]{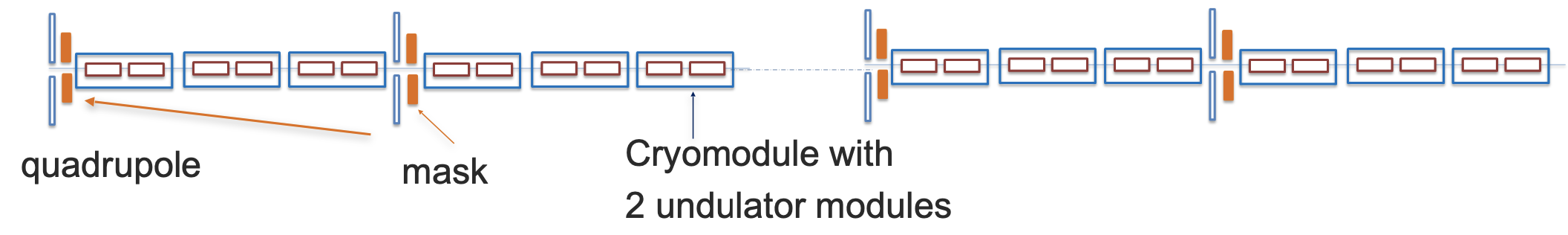} 
    \caption{Design of the mask system to protect the undulator walls, keeping the incident power of photons (including secondary particles) below the acceptable limit of 1 W/m \cite{Khaled}.}
    \label{fig:positron-masks}
\end{figure}

In addition a mask system, see Fig.~\ref{fig:positron-masks},  for the undulator walls has been designed, since the produced undulator photons are produced with an opening angle and some part of the photon beam will hit the wall. The acceptable level of the incident power at the wall has been assumed to be 1 W/m. The proposed masks have a length of 30 cm and 0.44 cm diameter. Several materials (Iron, Copper and Tungsten) have been studied. Such a mask system decreases the possible incident power on the undulator walls (including even secondary particles) by about two orders of magnitude, keeping it below the acceptable level even for the energy stage $\sqrt{s}=250$~GeV for the complete active undulator length, see Fig.~\ref{fig:sim-masks} (left panel).

\subsection{GigaZ option}
Electroweak precision observables (EWPO) $m_Z$, $m_W$, $\sin^2 \theta_W$ are crucial observables, in particular since the scale of new physics is still unknown. Even small traces from virtual effects of new physics can be deduced from deviations of EWPO measurements from their SM predictions.  
As well known from the experience at LEP and at SLC, ---where SLC achieved the best single measurement of $\sin^2 \theta_W$ via the left-right asymmetry $A_{\rm LR}$ compared to the forward-backward asymmetry $A_{\rm FB}$ at LEP although SLD had one order of magnitude less luminosity than LEP,--- it is substantial for electroweak precision measurements to run both with polarized beams and high luminosity at the Z-pole. 

\begin{figure}[H]
	\centering 
	\includegraphics[scale=.42]{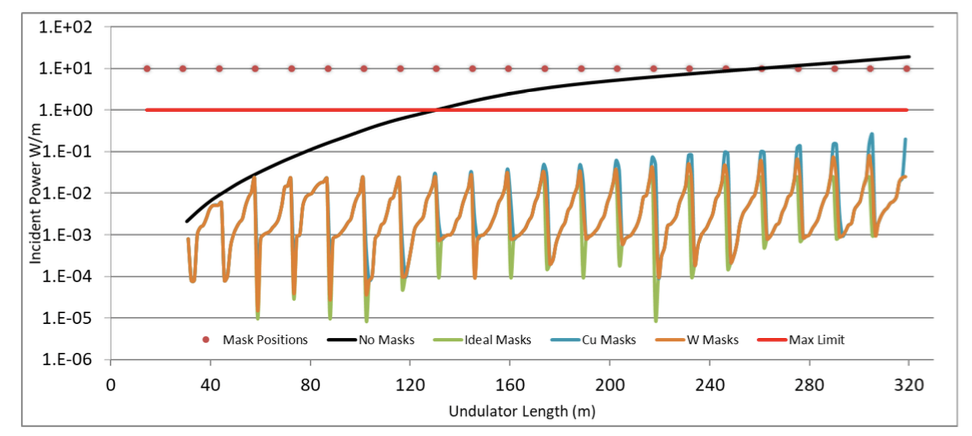}
	\includegraphics[scale=.3]{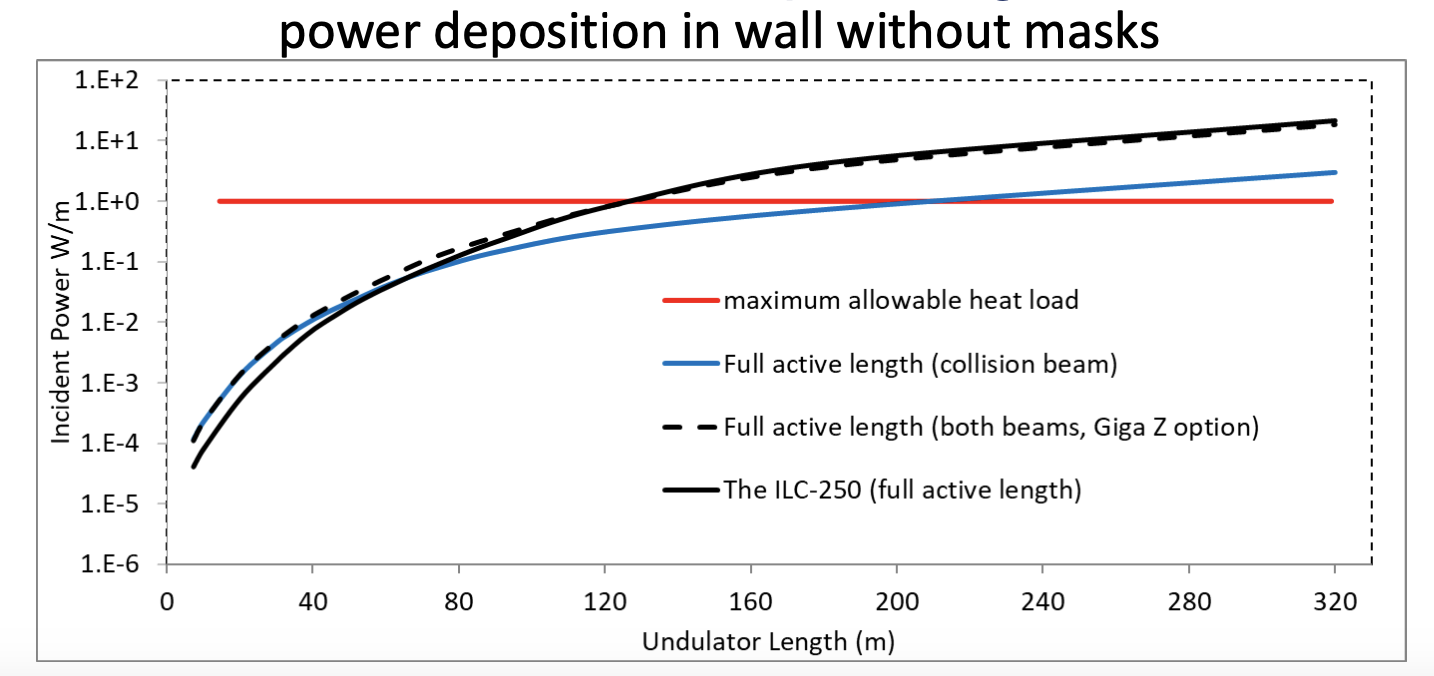}	 
    \caption{Impact of implementing the masks ( allowed limit of 1 W/m assumed (red line)) for protecting the undulator walls for the ILC250 (left panel) and comparing ILC250 and GigaZ option without masks (right panel)~\cite{Khaled}. Different materials iron, copper, tungsten have been used (left panel). Comparing the incident power for both options, ILC250 and GigaZ, indicates that the GigaZ leads to a comparable power deposition in the walls. Therefore the mask systems will be sufficient for both options\cite{Khaled}.}
    \label{fig:sim-masks}
\end{figure}

The GigaZ option at the ILC is foreseen to use a $3.7+3.7$Hz scheme\cite{Yokoya:2019rhx}, i.e. alternating an electron beam with 45.6 GeV for physics collisions and an electron beam with 125 GeV for the generation of positrons. Due to the lower electron energy the opening angle  of the produced photons in the undulator gets larger, i.e.\ more photons hit the undulator walls, however, with lower average energy compared to the 125-GeV electron beam.
Nevertheless, the deposited power in the undulator walls has to be checked for the GigaZ option as well.
Therefore in \cite{Khaled} both options have been calculated ---without using masks--- and as can be seen from Fig.~\ref{fig:sim-masks}(right panel), the 
incident power for the 3.7+3.7 G=Hz GigaZ option (dashed line) is always slightly lower/similar as for the ILC250 set-up. That means, applying the proposed mask scheme will also be sufficient for the GigaZ option.

\section{Positron Target Load for the undulator-based source} 
Due to the request for high luminosity the positron target load is in general for all high-energy lepton collider designs rather high.
However, using rotating wheels and therefore mitigating the target load on a specific position, leads to a viable design.
\subsection{Rotating wheel design simulations}
The current design  foresees that the photon beam hits a 1m diameter target wheel at the rim, spinning with 100 m/s tangential speed, so that the heat load
is well distributed: one pulse with 1312 (2625) bunches occupies about 7 (about 10) cm, i.e. only every 7-8 seconds the photon beam hits the same position at the target. For the ILC250 set up the first harmonic of the produced photons is at about 7.5 MeV and a target thickness of 7mm has been foreseen to optimize the $e^+$ yield in concordance with the acceptable deposition.

\begin{figure}[H]
	\centering
	\includegraphics[scale=.55]{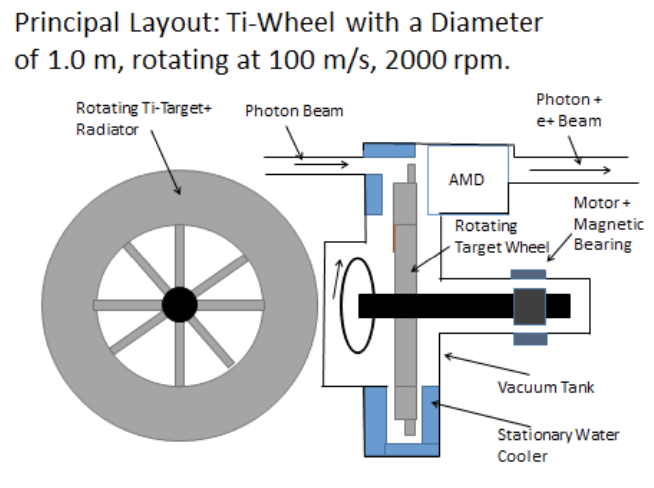}
	\hspace{1.2cm}
	\includegraphics[scale=.5]{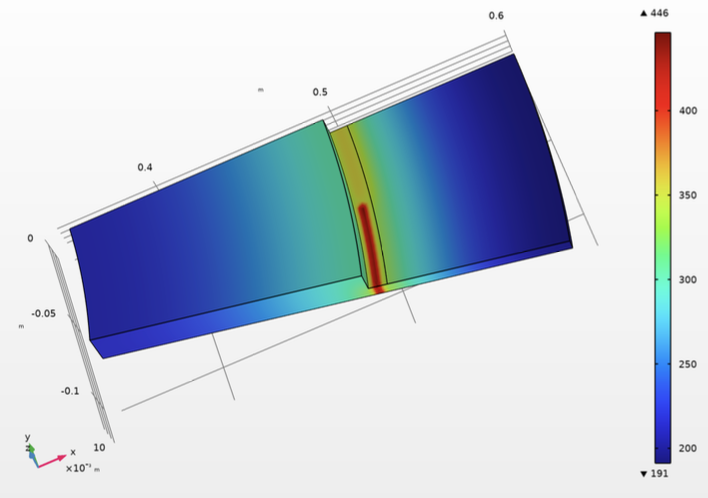}	 
    \caption{Left panel: Rotating Wheel, rotating in vacuum, supported by magnetic bearings and embedded in a stationary water cooler. Right panel: COMSOL simulations of the temperature load at the rotating target \cite{Samanwaya}.}
    \label{fig:rotating-wheel}
\end{figure}

Nevertheless the target has to be cooled and it has been studied that the cooling via radiation of such 
a fast spinning wheel is absolutely sufficient\cite{Riemann:2018gif}: exploiting fully the Boltzmann $T^4$-law, leading to a peak temperature  of about $550^0$C for the ILC250 option and to about $500^0$C for the GigaZ option, both cases with 1312 bunches/pulse. Even the luminosity update version
has already been studied and should be compatible.
Since radiation cooling is sufficient one has the option to use magnetic bearings for the rotation, see Fig.~\ref{fig:rotating-wheel}(left panel):
the rotation axis is floating in the magnetic field, the whole device is embedded in vacuum surrounded by a stationary water cooler.

New simulations, Fig.~\ref{fig:rotating-wheel}(right panel) have been performed and are still ongoing~\cite{Samanwaya} for the rotating wheel design in combination with the optic matching device, the pulsed solenoid, see sect.~\ref{ref:OMD}. Technical specifications for the specific ILC application are currently updated~\cite{Sievers}.

\subsection{Target load tests}
Detailed target material tests at the Mainz Microtron (MAMI) have been performed since several years~\cite{Ushakov:2017dha,Heil:2017ump,Staufenbiel:2014hia, Riemann:2012zz}. The electron beam at MAMI generates a cyclic load with at least similar but often a higher peak energy deposition density (PEDD)
than expected at the ILC. Several Ti-alloys with different target thickness, isolator and thermocouple set-ups have been studied  and have been analyzed with both laser scanning methods but also with a synchrotron diffraction methods, analyzing the different $\alpha$- and $\beta$-phases of the Ti-alloy, 
\cite{Tim-Arbeiten,Tim-proc}. In order to differentiate whether structural material changes originate from radiation or from temperature variations, the  targets have also been set in both fast and cyclic thermic stress in the range of $400^0$-$800^0$C via a dilatometer and compared with the radiated targets. There occur transitions between the $\alpha$- and $\beta$-phase of Ti-alloys, but the overall result is that the foreseen ILC positron source target material, Ti-6Al-4V will stand the load. More details see~\cite{Tim-proc,Tim-Arbeiten}.

\section{Optic matching devices}\label{ref:OMD}
%
A special feature of the undulator driven positron source is the 1\,ms long photon pulse, incident on the rotating wheel. This is much longer than usually used for conventional positron sources which usually have micro-second pulses and where Flux concentrators (FC) are convenient.  However, due to the time varying skin effect in these FCs, when driven with ms-pulses, the magnetic field will strongly vary during the beam pulse which has an impact on the luminosity and is not favourable for high precision measurements.

Therefore a pulsed solenoid is currently designed as OMD for the ILC positron source and prototyping work has started recently, see Sect.\ref{ref:ps}.
Furthermore also a plasma lens is discussed for future application at the ILC positron source and first steps towards a proof-of-principle experiment with a
small prototyp lens have been done, see Sect.\ref{ref:pl}.

\subsection{Pulsed solenoid}
\label{ref:ps}
Pulsed solenoids as  OMD have been used in the past, for instance, for the positron source of CERN-LEP.  
In general a pulsed solenoid provides a stable and reproducible focus, offers a high magnetic flux density and is compatible with long pulse durations.

\begin{figure}[H]
	\centering
	\includegraphics[scale=.6]{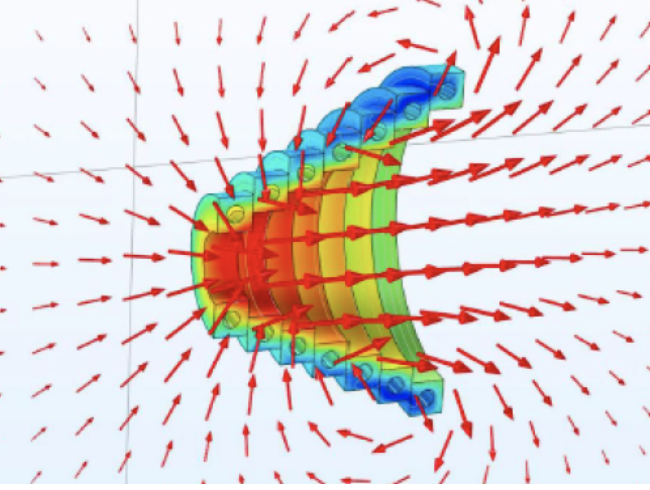} 
		\hspace{1.2cm}
	\includegraphics[scale=.6]{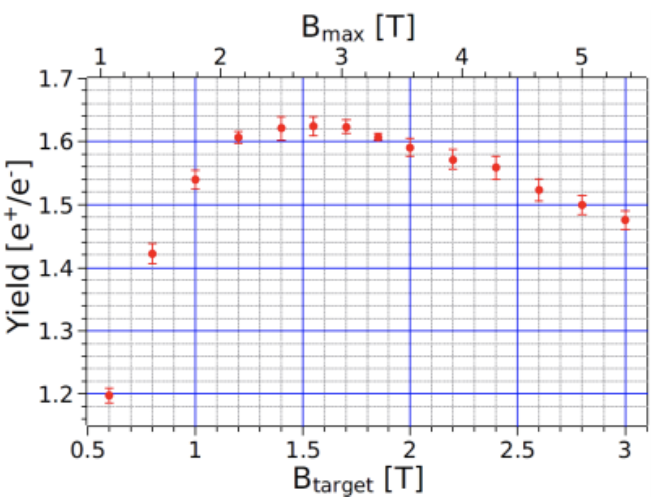}
    \caption{Left panel: Pulsed solenoid fields, half-sine current pulse with 4 ms and a peak current of 50 kA and peak field of 5.3 T. -- Surface: Magnetic flux density norm (T), Arrow volume: Magnetic flux density (spatial frame), generated with COMSOL-Multiphysics Code \cite{Mentink};
    Right panel: Expected positron yield depending on the field $B_{target}$ at the target exit and the maximum field $B_{max}$ \cite{Riemann:2020ytg}.}
    \label{fig:pulsed-solenoid}
\end{figure}

For the ILC positron source, a basically simple solenoid, wound in a conical shape is considered , see Fig.~\ref{fig:pulsed-solenoid} (left panel), and the expected yield in dependence of the magnetic field can be seen in Fig.~\ref{fig:pulsed-solenoid} (right panel).
In order to achieve multi-Tesla fields, for instance, 5 Tesla, peak currents of about 50 kA or above are required. Clearly, with such high currents, the solenoid cannot be driven in a d.c. mode, but has to be pulsed. Considering a Cu-conductor of about 1\,cm$\times$1\,cm,  the pulse duration has to be chosen long enough, so that at the peak of the half-sine pulse, eddy current have died out and a stable field over the duration of the beam pulse  of 1\,ms is achieved.  
Such a stability will be reached with a half sine current pulse  with a duration of about 4 ms, where the skin depth in Cu will be about 0.6\,cm, sufficiently larger than the average radius of the conductor  with a resistivity of about $2\times 10^{-8}\,\Omega$m.  This pulsing will still provide a large reduction of the electrical power consumption of the solenoid with a duty cycle of only 1\%, when driven at 5\,Hz. 

\begin{figure}[H]
	\centering
	\includegraphics[scale=.35]{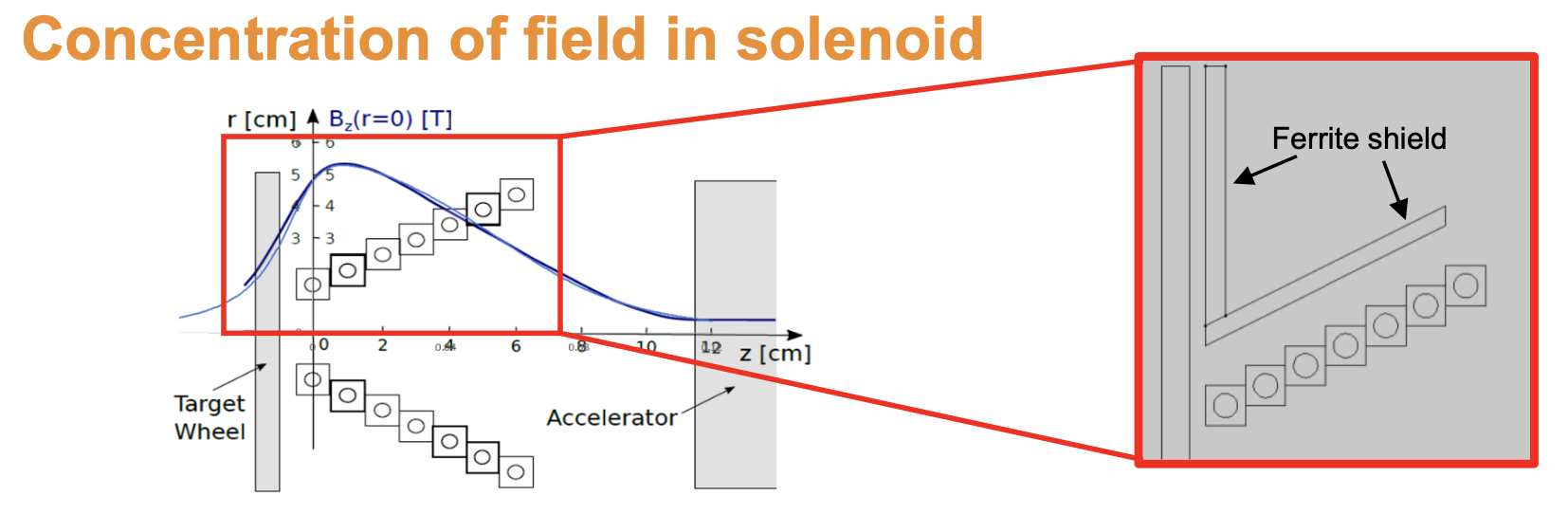} 
		\hspace{.2cm}
	\includegraphics[scale=.42]{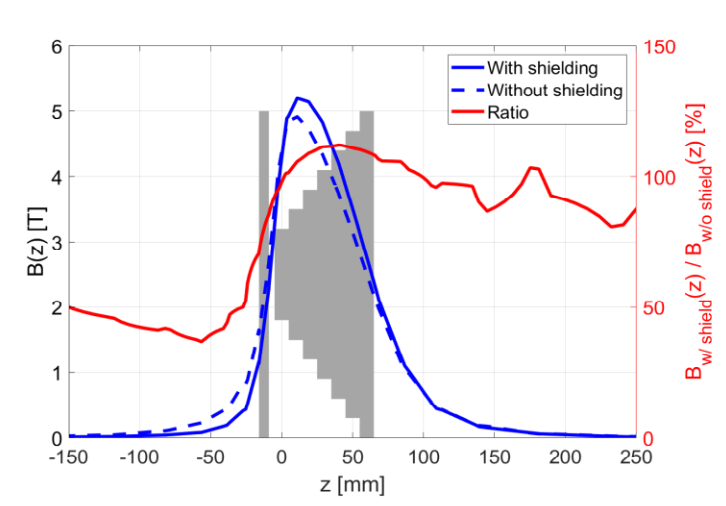}
    \caption{Left panel: Ferrite shielding inside the pulsed solenoid, 2D and 3D simulations with COMSOL including the rotating target (100 m/s) and a peak current of 45 kA \cite{Carmen}. Right panel: Resulting magnetic field with and without shielding. The induced heat  as well as the peak force are reduced by more than a factor two with shielding, the peak field is slightly increased by $B(z)~10\%$.  \cite{Carmen}}
    \label{fig:shielding}
\end{figure}

The average heat load on the target during pulse and flat-top is about 73~W+711~W,
and the peak force on the rotating wheel is about 612 N~\cite{Carmen} and a ferrite shielding will be implemented~\cite{Carmen}, see Fig.~\ref{fig:shielding}(left panel). Both 2d and 3D detailed simulations have been performed with COMSOL w/wo the moving Titanium plate with 100 m/s and a peak current of 45 kA. The shielding reduces slightly the peak magnetic flux, by less than 10\%. However, it reduces significantly the induced heat load by more than a factor two, 31~W+298~W, so that the peak force on the rotating wheel is reduced to 263~N.  The peak magnetic field $B(z)$ increases by about 10\%, see Fig,~\ref{fig:shielding}(right panel).

The yield of the undulator-based source with the pulsed solenoid as OMD has been simulated up to the damping ring (DR) and a significant yield increase compared to a quarter-wave-transformer (QWT) as OMD has been achieved, see Tab.\ref{tab:yield} \cite{Fukuda,Carmen}.

\begin{table}
\centering
\begin{tabular}{|c|c|c|}
\hline
 & \multicolumn{2}{|c|}{Positron Yield}\\
  OMD & $@$Capture ($|z|<7$mm) & $@$Damping Ring\\\hline
 Quarter-Wave-Transformer & 1.07&  $\sim$1.1\\
 Pulsed Solenoid without shielding & 1.81 & 1.91 \\
 Pulsed Solenoid with shielding & 1.64 & 1.74\\ \hline
   \end{tabular}
\caption{Expected positron yield \cite{Fukuda} at the undulator-based positron source for different optic matching devices, the Quarter-Wave transformer (QWT) and the Pulsed Solenoid w/wo the ferrite shielding \cite{Carmen}. \label{tab:yield}}
\end{table}

The stress in the coil of the pulsed solenoid has been simulated as well and a average power of about 10~kW is expected for a peak magnetic flux of 4.6~T.
The von-Mises stress in the coil amounts to about 570 MPa and provides  a conservative estimate on the expected stress. However, the
 soft tensile strength of copper is about 200 MPa. The actual stress in the coil is very sensitive to the exact shape.
 Therefore the mechanical design including the solenoid, the support structures and the connectors have to be iterated. Even in case the stresses get too high, one could find solutions, for instance, using multiple layers or reducing the peak current~\cite{Carmen,Formela:2022gco}. 
 
\begin{figure}[H]
	\centering
	\includegraphics[scale=.35]{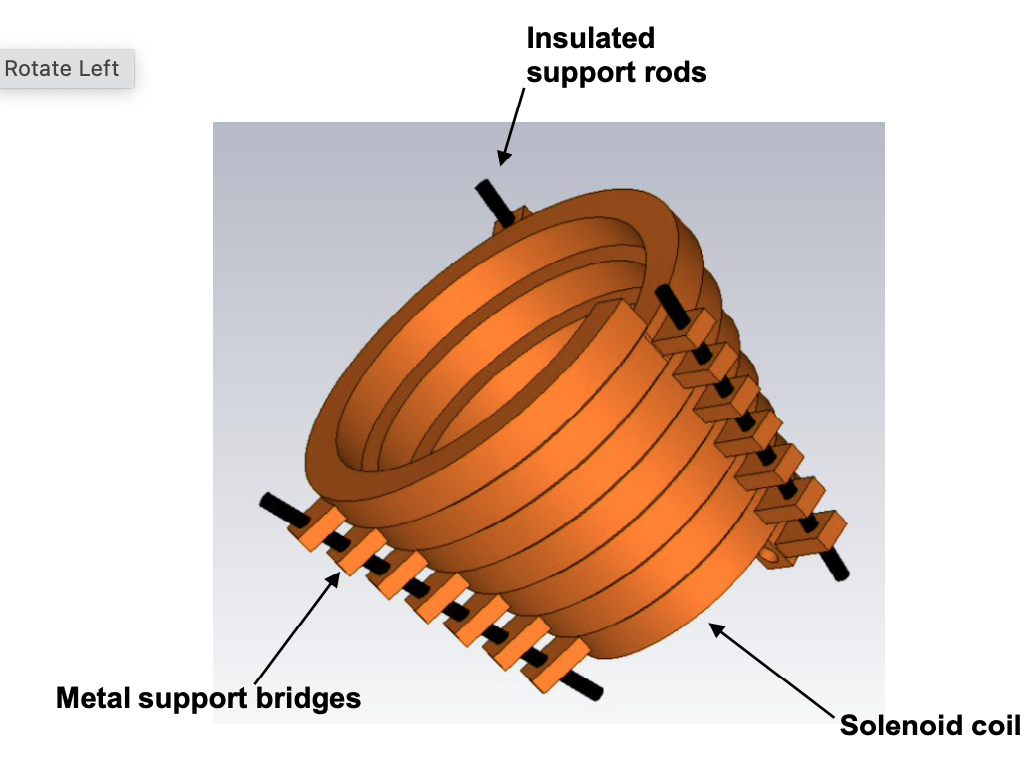} 
		\hspace{.2cm}
	\includegraphics[scale=.42]{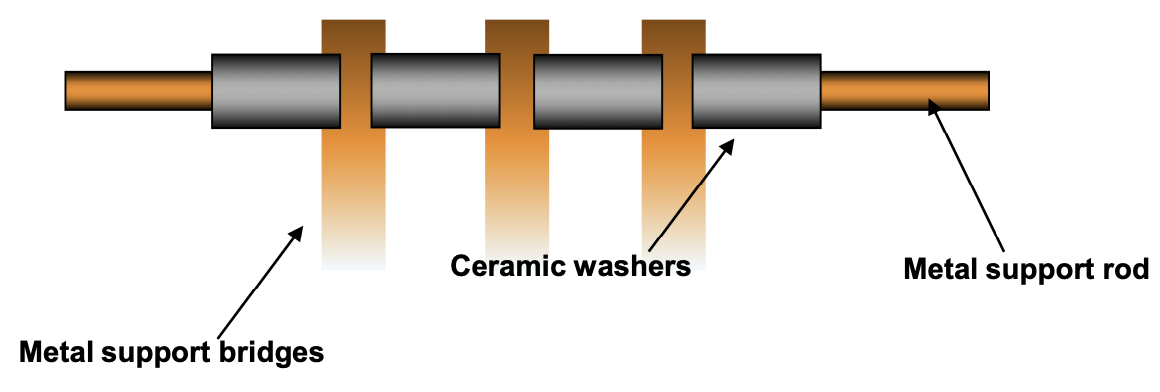}
    \caption{Left panel: Design of the tapered solenoid coil with 7  windings, metal support bridges and insulated support rods  \cite{Carmen-eps}. Right panel: layout of the prepared prototype \cite{Carmen-eps}. }
    \label{fig:design}
\end{figure}

 The global optimization is therefore still outstanding awaiting first results  from the prototype of the pulsed solenoid design.
Mechanical drawings for such a prototype have been done recently, see Fig.~\ref{fig:design}, consisting of a solenoid coil with tapered seven planar windings and interconnections. The conductor is cooled from inside. The coil  is hold via metal supporters, the rods are insulated from the bridges, i.e. the magnetic shielding is cut at the support locations, which might have some influence on the field, but the main shielding of the target is expected to be unaffected. First measurements of this prototype are envisaged for this year until next spring~\cite{Carmen,Carmen-eps}.

\subsection{Plasma lens}
\label{ref:pl}
\begin{figure}[H]
	\centering
	\includegraphics[scale=.5]{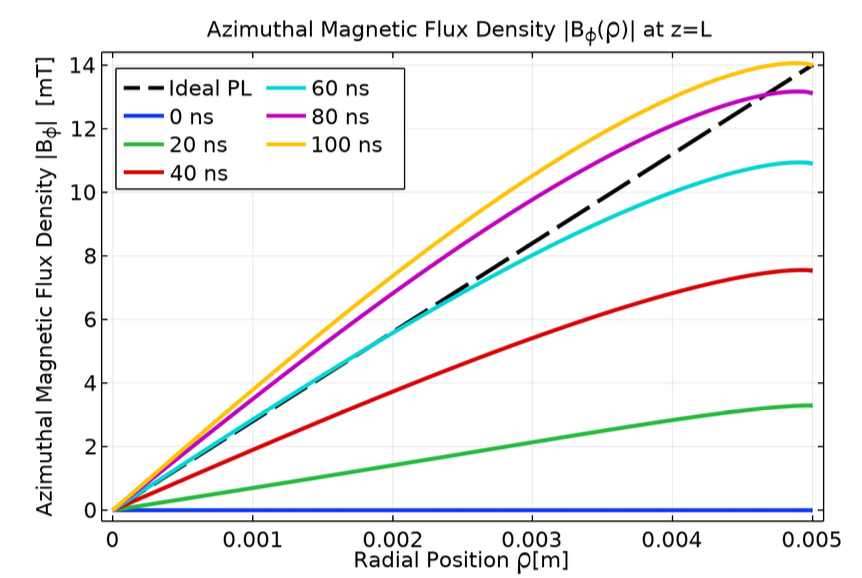} 
    \caption{Radial distribution of the azimuthal magnetic flux density at the exit of the active plasma lens: hydrodynamical simulations (solid lines) compared with  an ideal model with homogeneous current density (dashed line) \cite{Manuel-proc,ipac-23-manuel}.}
    \label{fig:sim-pl}
\end{figure}

\begin{figure}[H]
	\centering
	\includegraphics[scale=.5]{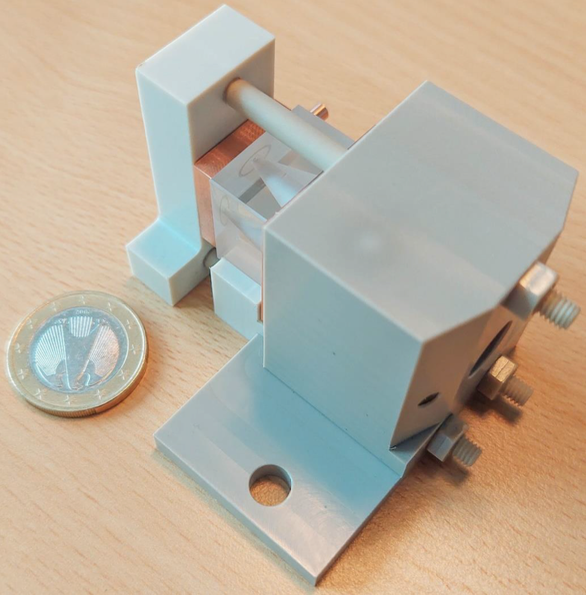} 
		\hspace{1.2cm}
	\includegraphics[scale=.7]{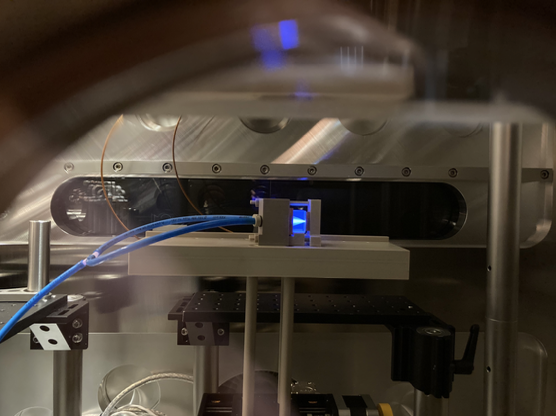}
    \caption{Left panel: Image of the prototype plasma lens set-up. The coin serves only for illustrating the proportions. Right panel: Produced plasma in the prototype lens \cite{Niclas-proc,ipac-23-niclas}.}
    \label{fig:plasma-lens}
\end{figure}

Using active plasma lenses as OMD is currently discussed as a novel application that has a  high potential for improving the yield and collecting the highly divergent positrons. A tapered, azimuthal magnetic flux is generated in the plasma causing radial forces on the traversing beam. Such a tapered plasma lens is a novelty and detailed hydronamical simulations, see Fig.\ref{fig:sim-pl}  and construction work, see Fig.\ref{fig:plasma-lens}(left panel), towards a downscaled prototype experiment have been made. More details can be found in \cite{ipac-23-manuel, ipac-23-niclas}.
Recently, the  first tests and measurements have been successfully started and a plasma has been generated inside the prototype, see Fig.\ref{fig:plasma-lens} (right panel). Measurements of the field as well as further tracking simulations are currently under work, more details
 see~\cite{Niclas-proc, Manuel-proc}.

\section{Conclusions}
The undulator-based positron source for the ILC is a mature design and offers in addition a polarized positron beam.
Simultaneously polarized beams offer several advantages, for instance, a higher effective polarization and luminosity, better statistics, control of the limiting systematics and additional, new observables ---as well as exploiting transversely-polarized beams--- for detecting new physics already at the first energy stage of $\sqrt{s}=250$~GeV. 

There exist several yers of experience with successful running of long undulators at the XFEL accelerator. The alignment is successfully controlled and stable long-time running has been achieved.
Nevertheless, further involved simulations for the ILC undulator have been performed, including non-ideal field distributions as well as a mask design set-up for protecting the undulator walls against incident photon energy deposition. Such a protection system is also applicable and sufficient for the ILC 3.7+3.7 Hz GigaZ option.

Involved target tests and analyses have been done. ILC targets with different thickness and set-ups have received a corresponding PEDD as expected during ILC running times via the electron beam at MAMI and the target materials have been analyzed with involved diffraction methods. Furthermore, pure thermic stress tests  with a dilatometer have been applied on ILC targets as well. The results show that the thermic stress is decisive but that the ILC targets are safe and will stand the load.
 
 In addition new detailed simulations exists for the optic matching device, the pulsed solenoid, in combination with the fast rotating wheel. With the pulsed solenoid the yield requirements will be safely matched. Currently mechanical drawings, engineering and tests for a prototype of the pulsed solenoid are ongoing. Also for new technology devices, plasma lenses as optic matching devices, a first downscaled prototype has been constructed and is currently being tested. Advanced hydrodynamical plasma simulations are progressing as well.

Altogether, the baseline ILC undulator-based source  is a mature design matching all ILC requirements and offering in addition polarized positrons, that are substantial to fulfill the physics expectations and to open possible windows to new physics areas.

A further interesting application is the adaption of the undulator-based positron source for the hybrid, asymmetric, linear Higgs factory
based on plasma-wakefield and radio-frequency acceleration (HALHF) concept, design considerations have been started.

\section*{Acknowledgements}
GMP acknowledges support by the Deutsche Forschungsgemeinschaft (DFG, German Research Foundation) under Germanys Excellence Strategy --EXC 2121 Quantum Universe-- 390833306.

\end{document}